\documentclass[Journal]{IEEEtran}

\usepackage{graphicx} 
\usepackage{subfigure}
\usepackage{epstopdf}
\usepackage[square,sort&compress,comma,numbers]{natbib}
\usepackage{color}
\usepackage{amsfonts}
\usepackage{latexsym}
\usepackage{amssymb}
\usepackage{amsmath}
\usepackage{fixltx2e}
\usepackage{multirow}
\usepackage{upgreek}
\usepackage{algorithm}
\usepackage{algpseudocode}
\usepackage{mathrsfs}
\usepackage{cancel}

\usepackage[table,xcdraw]{xcolor}
\IEEEoverridecommandlockouts
\begin{document}
\title{LightSleepNet: Design of a Personalized Portable Sleep Staging System Based on Single-Channel EEG}

\author{\IEEEauthorblockN
        {Yiqiao Liao\IEEEauthorrefmark{2},
        Chao Zhang\IEEEauthorrefmark{1},
        Milin Zhang\IEEEauthorrefmark{1},
        Zhihua Wang\IEEEauthorrefmark{2},
        Xiang Xie\IEEEauthorrefmark{2}
        }

\thanks{Yiqiao Liao, Zhihua Wang and Xiang Xie are with the Institute of Microelectronics, Tsinghua University, Beijing, China, 100084. }
\thanks{Chao Zhang and Milin Zhang are with the Department of Electronic Engineering, Tsinghua University, Beijing, China, 100084. Corresponding author e-mail: zhangmilin@tsinghua.edu.cn}
\thanks{This work is supported in part by the National Key Research and Development Program of China (No.2018YFB220200*), in part by the Beijing Innovation Center for Future Chip,  in part by the Beijing National Research Center for Information Science and Technology.}
\thanks{Digital Object Identifier}
}

\maketitle
\begin{abstract}
This paper proposed LightSleepNet - a light-weight, 1-d Convolutional Neural Network (CNN) based personalized architecture for real-time sleep staging, which can be implemented on various mobile platforms with limited hardware resources. The proposed architecture only requires an input of 30s single-channel EEG signal for the classification.
Two residual blocks consisting of group 1-d convolution are used instead of the traditional convolution layers to remove the redundancy in the CNN.
Channel shuffles are inserted into each convolution layer to improve the accuracy. In order to avoid over-fitting to the training set, a Global Average Pooling (GAP) layer is used to replace the fully connected layer, which further reduces the total number of the model parameters significantly.
A personalized algorithm combining Adaptive Batch Normalization (AdaBN) and gradient re-weighting is proposed for unsupervised domain adaptation. A higher priority is given to examples that are easy to transfer to the new subject, and the algorithm could be personalized for new subjects without re-training.
Experimental results show a state-of-the-art overall accuracy of 83.8\% with only 45.76 Million Floating-point Operations per Second (MFLOPs) computation and 43.08 K parameters.
\end{abstract}


\IEEEpeerreviewmaketitle

\section{Introduction}

Sleep is important for humans to keep the nervous system functioning well. 
Unfortunately, more than 20 percent of the adult population are suffering from various sleep disorders \cite{population-sleep-disorder}.  
Sleep staging can be applied for the diagnosis and treatment of sleep disorders \cite{pachori2020biomedical}.
Polysomnography (PSG) based sleep staging is widely used in clinical practice.
It is the golden standard as experts label the sleep stages according to the recorded Electroencephalography (EEG), Electrooculography (EOG), Electromyography (EMG) and Electrocardiogram (ECG). 
However, it is difficult to apply at home due to the complex operation process of sleep staging. 
In addition, the requirement for real-time sleep staging is emerging while exploring effective methods to improve the sleep quality, such as sounds, lights and electrical stimulation \cite{leminen2017enhanced}. 
As a result of the development of the wearable personal health monitoring devices in recent years, a long-term, real-time, high-precision sleep staging algorithm is required for implementation in various portable devices \cite{portable-sleep}.

Hardware-friendly algorithms with low computational complexity have been explored to fit sleep staging process in wearable devices. 
Traditional machine learning based methods \cite{sharma2017automatic,imtiaz2017ultralow,bajaj2013automatic}, such as decision tree  \cite{imtiaz2017ultralow} or support vector machine \cite{bajaj2013automatic} based classifiers, can be implemented in wearable or mobile devices, but suffer from low accuracy, usually lower than 80\% \cite{imtiaz2017ultralow}.
Deep learning based algorithms have been widely applied to improve the performance of biomedical signals (e.g. EEG) processing in recent years \cite{srirangan2019time, das2010handbook}. 
A considerable amount of literature \cite{supratak2017deepsleepnet,phan2019seqsleepnet,liao20Trifeature,phan2019towards,chambon2018deep,phan2018dnn,phan2018joint,koushik2019real,chang2019ultra} has been published on automatic sleep staging based on deep learning. 
The SeqSleepNet \cite{phan2019seqsleepnet} achieved an accuracy of near 90\%. 
However, it suffers from a difficult compromise between observation latency and computational complexity, since it requires ten raw EEG epochs together as the input. 
Time-Distributed Deep CNN has been applied to fit the requirement of real-time processing \cite{koushik2019real}, showing a promising result but the computational complexity is high. 

The individual differences raise another challenge for automatic sleep staging system.
Features extracted from the EEG signals distribute differently between the training set and test set, which makes the algorithms mentioned above unreliable for new subjects. 
Traditional solutions require fine-tuning using labeled data from the target subject for personalization \cite{attaran2018embedded}. 
However, professional knowledge is required for raw data labeling, which is unavailable at home.
\cite{liao20Trifeature} proposed to solve this problem with adversarial training, but there was sleep information lost in the training process, which resulted in a low accuracy.
\cite{khalighi2012adaptive} proposed to apply weighted kernel logistic regression for handcrafted feature extraction. 
However, a re-training of the network is required for most domain adaptation methods, which is both hardware hungry and power hungry.

This paper proposes a light-weight personalized sleep staging algorithm, which is denoted as LightSleepNet. 
The proposed architecture is suitable for implementation on various mobile platforms for real-time processing.
To reduce the negative influence of individual differences, unlabeled data can be used to personalize for new subject without re-training.
A 1-d CNN is designed for the feature extraction from 30s single-channel EEG epochs. 
In order to remove the redundancy in the CNN, residual blocks consisting of group 1-d convolution bring a dramatic reduction on the complexity of the network. 
Channel shuffles are inserted into each convolution layer to improve the accuracy. 
In order to avoid over-fitting to the training set, a Global Average Pooling (GAP) layer is used to replace the fully connected layer, which further reduces the total number of the model parameters by 12 times.

In order to further improve the algorithm robustness to individual differences without a significant increase of the complexity, a personalized unsupervised domain adaptation algorithm is proposed. 
Inspired by \cite{li2018gradient}, the gradient contribution of different samples could be re-weighted to improve the generalization. 
In our proposed work, a higher priority will be given to those examples that are easier to transfer to new subjects based on the gradient re-weighting in training.
The proposed Adaptive Batch Normalization (AdaBN) \cite{li2018adaptive} based method is designed for subject-specific adaptation,
which normalizes the intermediate output of CNN from training set and the data from the new subject to a similar distribution.

The rest of the paper is organized as follows. 
Section II introduces the proposed light-weight architecture and details of the proposed low complexity solution to improve the algorithm robustness to individual differences. 
Section III illustrates the experimental results, while Section IV concludes the work.

\section{Architecture of the Proposed LightSleepNet}
\subsection{The Process of the LightSleepNet}
\begin{figure}[htb]
    \centering
    \subfigure[]{
    \includegraphics[width=0.4\textwidth]{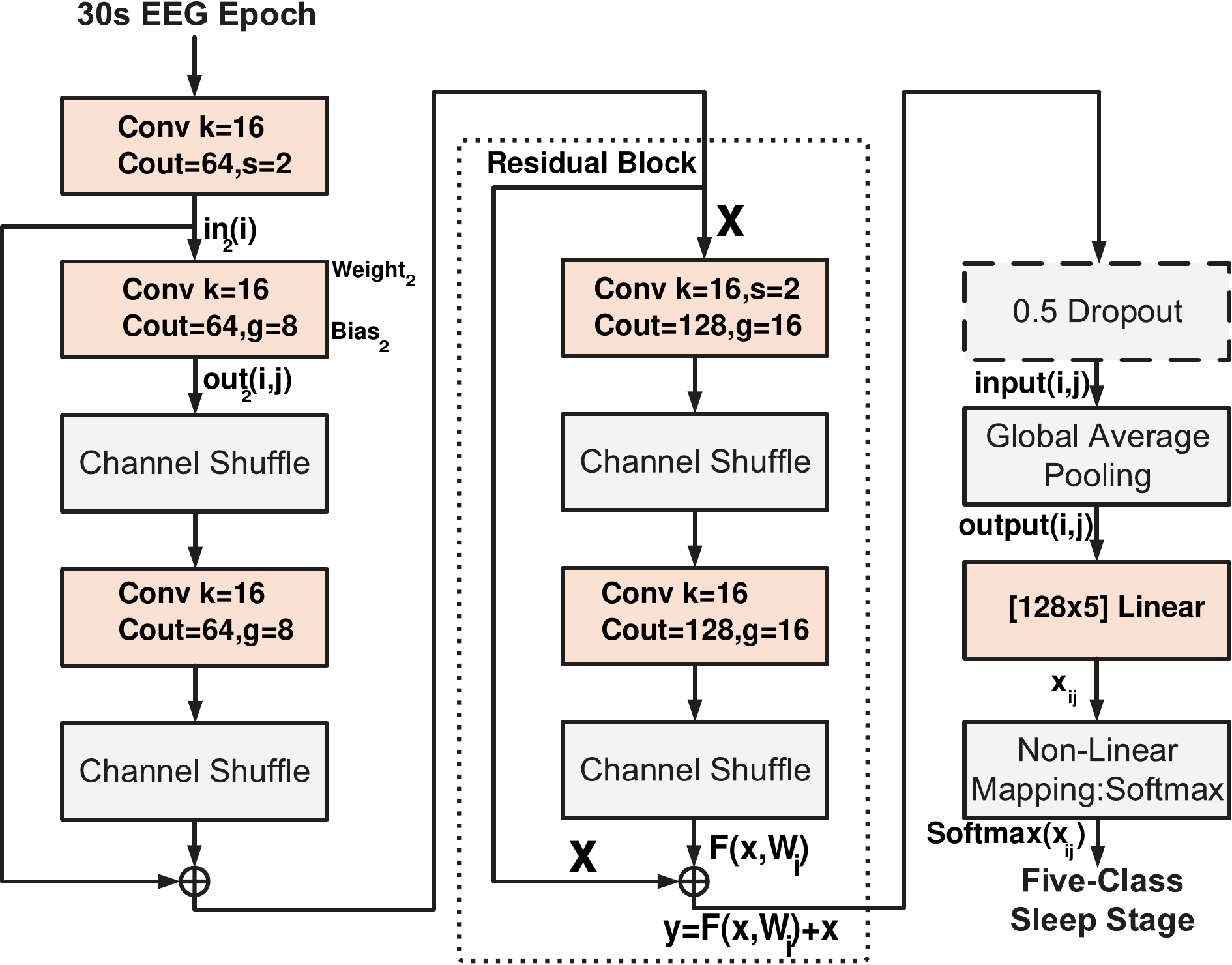}
    }
    \quad
    \subfigure[]{
    \includegraphics[width=0.4\textwidth]{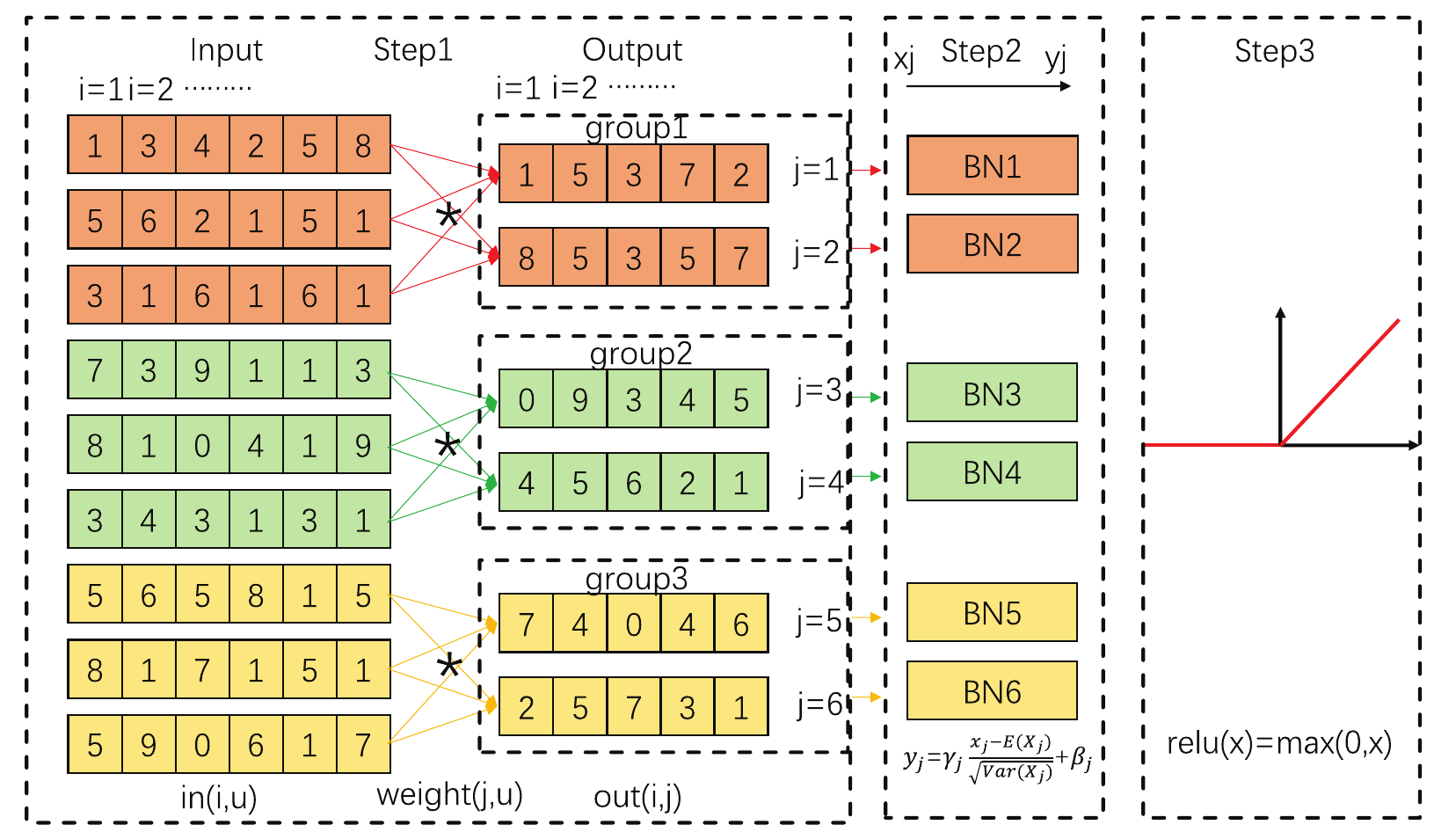}
    }
    \caption{(A) The overall architecture of LightSleepNet. $C_{out}$ and $C_{in}$ are the numbers of feature channels in output and input respectively. $k$ is the kernel size of convolution. $g$ is the number of group for convolution. $s$ is the stride with a default value of 1. (B) Three steps in 1-d convolutional layer (the orange blocks in (A)). The vectors in step 1 example the input and output of the 1-d group convolution.}
    \label{fig:lightsleepnet}
\end{figure}

\figurename \ref{fig:lightsleepnet}A illustrates the proposed LightSleepNet. 
It consists of five 1-d convolutional layers (as illustrated in the orange blocks) and one GAP layer to lower the dimension of the feature map as well as to reduce the workload for classification. 

The input is a 30s single-channel EEG epoch, which is denoted as $X_i$. The 1-d CNN is built with residual blocks consisting of 1-d group convolution. As illustrated in \figurename \ref{fig:lightsleepnet}B, there are three steps in each 1-d convolutional layer:
\begin{itemize}
\item [1)]

1-D group convolution with its filters:

As the step 1 of \figurename \ref{fig:lightsleepnet}B shows, the j-th channel output feature for the i-th sample can be calculated as follow:

\begin{equation}
\label{eq:cnn}
\begin{aligned}
out_n(i,j)=bias_n(j)+\sum_{u}weight_n(j,u)*in_n(i,u)
\end{aligned}
\end{equation}
where $j=[1,..., C_{out}]$.
$*$ is the convolution operation and u is the corresponding index of input channel belonging to the same group with the j-th output channel. $n=[1,2,...,5]$ is the index of this convolution layer.

\item [2)]

adaptive batch normalization:

BN layers transform the features $x_j$ into $y_j$ as
\begin{equation}
\label{eq:bn}
\begin{aligned}
 {\rm{y}}_{\rm{j}} = {\gamma _j}\frac{{{x_j} - E({X_j})}}{{\sqrt {Var({X_j})} }} + {\beta _j}
\end{aligned}
\end{equation}
where $x_j$ and $y_j$ are the input and output scalars of the BN, respectively, with $j=[1,...,U]$. $U$ is the feature dimension. $X_j\in \Re^{N} $ is the j-th column of the input feature. $N$ is the size of a batch,  chosen as 40 in this paper. $\gamma_j$ and $\beta_j$ are the training parameters.
\item [3)]
rectified linear unit (ReLU) activations: relu(x)=max(0,x)
\end{itemize}

A channel shuffle is inserted after each 1-d convolutional layer to improve the information independence introduced by group convolution. The channel dimension of the output from group convolution is reshaped into $(g,\frac{C_{in}}{g})$. It is then transposed to $(\frac{C_{in}}{g},g)$, before flattened as the input of next layer.
In order to assist the network training with multi-scale information, residual connection is inserted between every two convolution layers to form a residual block. The input $x$ is mapped by the residual block through the function $\mathscr{F}(*)$ as
\begin{equation}
\label{eq:res}
\begin{aligned}
y=\mathscr{F}(x,{W_i})+x
\end{aligned}
\end{equation}
where $W_i$ are the parameters of the residual block.

A Dropout layer is applied after the residual blocks for regularization with a 50\% probability randomly setting some of the input tensor as zero.
A GAP layer is inserted after the Dropout layer.
A smaller fully connected layer with a dimension of $128\times 5$ is applied to the output of the GAP layer. 
A Softmax layer is applied to the output of the fully connected layer for classification.
There are five optional outputs, Wake, N1, N2, N3 and REM, according to the sleep staging definition by the American Academy of Sleep Medicine (AASM).

\subsection{Training Process Design}
\begin{figure}[htbp]
  \centering
    \begin{minipage}{0.24\textwidth}
        \begin{center}
            \includegraphics[width=\textwidth]{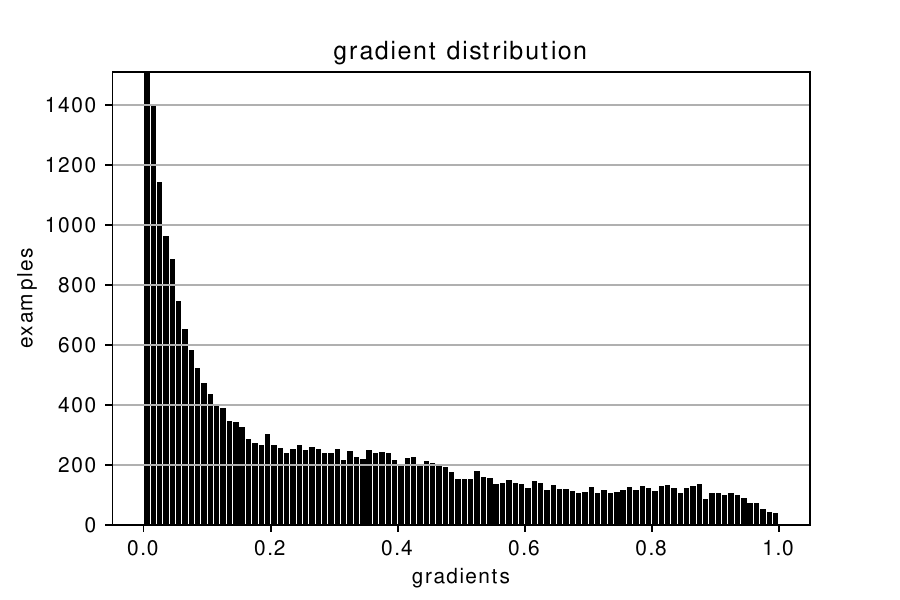}\\
            \scriptsize{(A)}\\
        \end{center}
    \end{minipage}
    \begin{minipage}{0.24\textwidth}
        \begin{center}
            \includegraphics[width=\textwidth]{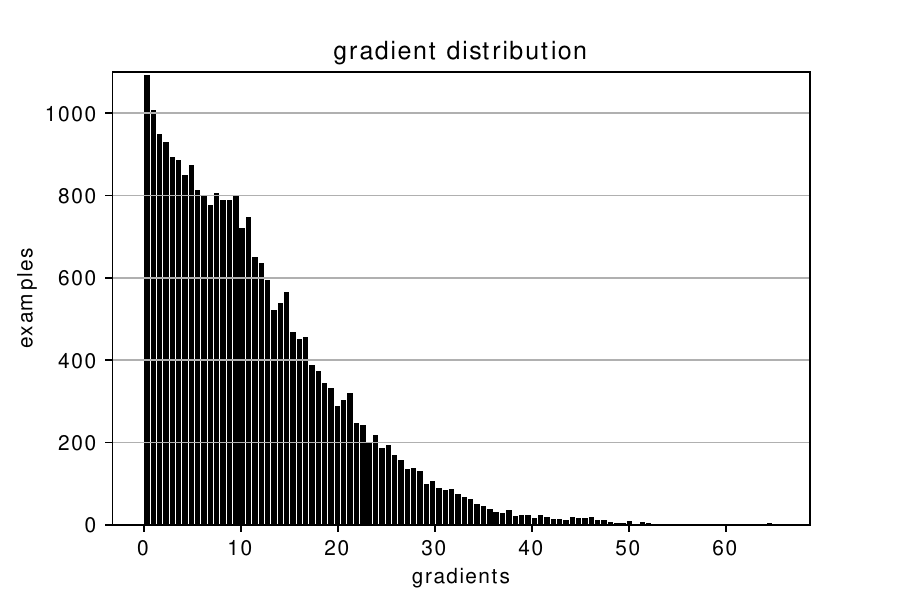}\\
            \scriptsize{(B)}\\
        \end{center}
    \end{minipage}
  \caption{(A) Gradient Distributed Before Weighting (B) Gradient Distributed After Weighting.}
  \label{fig:ghm}
\end{figure}

The traditional cross entropy loss imposes equal importance for different samples, whereas every sample does not contribute the same for the generalization.
There are samples hard to learn, which is usually a noise in the EEG signal and has lesser contribution to the generalization with uncertain predictions (i.e. large gradients). Those samples may deteriorate the training process. 
There are also samples easy to learn, which is easy to transfer to new subject with confident predictions (i.e. small gradients), contributing more to the generalization. 
We quantify the difficulty of samples by the norm of gradient and suppress those gradients from noise samples with lower weight, while giving high priority to those gradients from easy-to-transfer samples.
\figurename \ref{fig:ghm}A illustrates the gradient distribution for sleep staging. Gradient density is used to represent the number of samples within a specific gradient range. It is noted that there is a high gradient density in easy-to-transfer samples and low density in noise samples. We could give high priority to those samples with high gradient density.
We propose to replace the CrossEntropy loss in our model with a modified version to apply the re-weighting:
\begin{equation}\label{eq:ghm}
\begin{aligned}
L_{weighted}=\frac{1}{N}\sum_{i=1}^{N}\beta _i L_{CE}^{i}
\end{aligned}
\end{equation}
where $L_{CE}^{i}$ is the Cross-Entropy loss for the i-th sample. $N$ is the number of samples in one batch. $\beta _i$ is the weight of gradient for the i-th sample.
\begin{equation}\label{eq:ghm_beta}
\begin{aligned}
\beta _i=\frac{GD(g_i)}{N}
\end{aligned}
\end{equation}
where $g_i$ is the gradient of the i-th sample and $GD(g_i)$ is the gradient density of gradient $g_i$.

After the re-weighting, the gradient distribution is illustrated in \figurename \ref{fig:ghm}B, in which those samples easy-to-transfer initially focused on low gradient move right whereas those noise samples move left, which increases the number of samples with moderate gradient in the center region of the distribution. As the gradients are amplified for almost sixty times, a learning rate scheduler is applied for the adaptive learning rate adjustment.

\subsection{The AdaBN based Personalized Adaptation}
In order to improve the personalized adaptation, domain adaptation with unlabeled data from the new subject was applied. In the proposed work, AdaBN was used for subject-specific adaptation. 
AdaBN normalized features from training set and from the new subject to a similar distribution with zero bias and one variance without additional training cost.
The AdaBN is applied after every convolution layer.

The implementation of BN improves the similarity of the input distribution of each layer. The efficiency and robustness of the training process are improved as well. 
However, the data distribution and statistic values are different between the training and the test set when there is a significant domain shift.
The normalization process is unable to normalize these two datasets into a similar distribution since the statistic values in the target domain are not used. 
Compared to BN, it is better to use AdaBN when there are different statistic values between training stage and test stage, since the data distribution of the target domain is taken into consideration by the AdaBN.
With unlabeled data from a new subject, the batch statistic values are calculated offline or online based on forward propagation. With those statistic values, the feature distribution in new subjects would also be normalized to a similar distribution with zero bias and one variance. 
Each layer receives a similar input distribution for both data originated from the training set and a new subject. 
As a result, the domain shift between different subjects is alleviated with personalized adaptation. The process of training and testing with personalized adaptation is described in Algorithm \ref{alg:Training}.

\begin{algorithm}[htb]
  \small
  \caption{The process of training and testing with personalized adaptation.}
 \label{alg:Training}
 \begin{algorithmic}[1]
   \Require
     The set of input EEG epochs of source domain, $X_n^s$;
     The corresponding sleep stages labels of source domain, $Y_n^s$;
     The set of unlabeled EEG epochs of target domain, $X_n^t$;
   \Ensure
     The corresponding sleep stages labels of target domain, $Y_n^t$; \\
   \emph{\textbf{Training}}:
   \For{$epoch<100$}
   \State Forward Propagation to get prediction $\hat{Y}_{n}^{s}$ for the EEG input $X_n^s$ from source domain;
   \State Calculating the CrossEntropy Loss with function $L_{CE}=-\frac{1}{N}\sum_{j=1}^{N}\sum_{i=1}^{5} Y_{ji}^s log(\hat{Y}_{ji}^{s})$;
   \State Calculating the reweighting loss with weights corresponding to samples $L_{weighted}=\sum_{i=1}^{N}\frac{GD(g_i)}{N^2} L_{CE}^{i}$;
   \State Backward Propagation to update the model;
   \EndFor \\
   \emph{\textbf{Testing}}:
   \For{all neuron j in DNN}
   \State Calculating neuron responses $x_j$ on all EEG signals $X_n^t$ of target domain;
   \State Update the mean and variance of the target domain for that neuron using online algorithm:
   $\mu _j^t = {\rm E}(x_j^t),\sigma _j^t = \sqrt {Var(x_j^t)}$
   \EndFor
   \For{all neuron j in DNN}
   \State Calculating BN output on all EEG signals $X_n^t$ of target domain for neuron j:
   ${\rm{y}}_{\rm{j}} = {\gamma _j}\frac{{{x_j} - E({X_j})}}{{\sqrt {Var({X_j})} }} + {\beta _j}$
   \EndFor
   \State Forward Propagation to get prediction $Y_n^t$ for the EEG input $X_n^t$ from target domain using the BN output calculating above;
   \State
   \Return $Y_n^t$;
 \end{algorithmic}
\end{algorithm}

\subsection{The Computational Complexity of the LightSleepNet}
\begin{figure}[htb]
\centering
\includegraphics[width=0.35\textwidth]{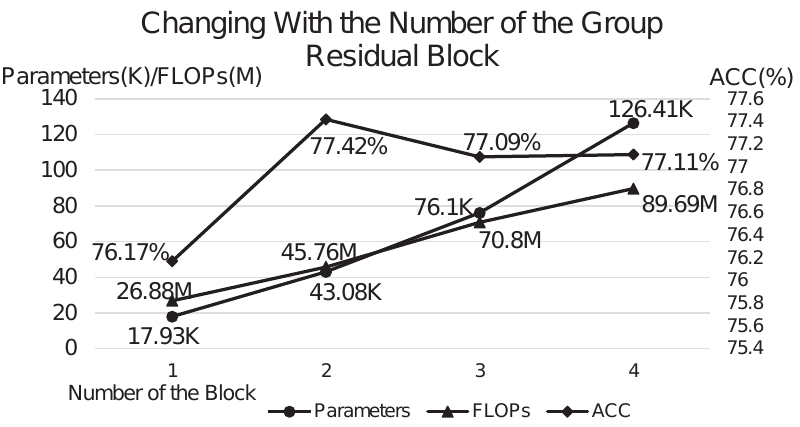}
\caption{Performance Changing With The Number Of Block}
\label{fig:numberofblock}
\end{figure}

Figure \ref{fig:numberofblock} shows that there is a trade-off between the accuracy (ACC) and the complexity of the network. The best performance is achieved with a residual block number of 2. 

Table \ref{tb:abalation} compares the ACC and computational complexity between different cases. 
It shows the improvement by applying group convolution, GAP layer, residual blocks and channel shuffles. In the proposed work, the number of parameters of one group convolution layer can be calculated as:
\begin{equation}
\label{eq:flops}
\begin{aligned}
(k*\frac{C_{in}}{g}*C_{out}+C_{out}) \propto \frac{1}{g}
\end{aligned}
\end{equation}
As a result, the number of parameters could be reduced by $g$ times, where $g$ is the number of groups.
The FLOPs of that layer can be calculated as:
\begin{equation}
\label{eq:flops_group}
\begin{aligned}
(k*\frac{C_{in}}{g}*C_{out}+C_{out})*\frac{L_{out}}{s}  \propto \frac{1}{g*s}
\end{aligned}
\end{equation}
where $L_{out}$ is the length of the output feature. $s$ is the stride size and $k$ is the kernel size.
According to eq.(\ref{eq:flops_group}), the FLOPs could also be reduced by $g$ times. As a result, the group convolution brings a 12 times reduction with a higher accuracy.

For the fully connected layer, the number of parameters and FLOPs can be calculated as:
\begin{equation}
\begin{aligned}
N_{in}*N_{out}+N_{out}
\end{aligned}
\end{equation}
where $N_{in}$ and $N_{out}$ are the counts of the input and output features, respectively. An $N_{out}=5$ is used for the sleep staging in the proposed work. The using of the GAP layer reduced the value of $N_{in}$ from 96256 to 128. As a result, a number of 480k parameters reduction and an improve of 3.24\% accuracy improvement are achieved. The introduce of channel shuffle features a 1.25\% improvement of the accuracy.

\begin{table}[!ht]
\caption{Comparison of the ACC and number of parameters between the proposed work under different cases}
\label{tb:abalation}
\begin{center}
\scriptsize
\begin{tabular}{c|c|c|c}
\hline\hline
Methods& Parameters &FLOPs& ACC \\
\hline
\hline
LightSleepNet&43.08K&45.76M & 77.42\%\\
\hline
Group Residual Block*1+GAP &17.93K&26.88M & 76.17\%\\
\hline
Group Residual Block*2+GAP &43.08K&45.76M & 77.42\%\\
\hline
Group Residual Block*3+GAP &76.10K&70.80M & 77.09\%\\
\hline
Group Residual Block*4+GAP &126.41K &89.69M& 77.11\%\\
\hline
Traditional Residual Block*2+GAP &526.4K&496.45M& 75.84\%\\
\hline
Group Residual Block*2 &523.72K &46.24M& 71.14\%\\
\hline
Group Block*2+GAP &43.08K & 45.76M&74.35\%\\
\hline
LightSleepNet Without Shuffle &43.08K & 45.76M&76.17\%\\
\hline\hline
\end{tabular}
\end{center}
\end{table}

It is noted that in \figurename \ref{fig:numberofblock}, the accuracy first increases with the model depth and then decreases due to over-fitting.
In order to deal with the over-fitting issue, dropout layer, regularization term and unlabeled data are used.
In addition, a group convolution and GAP layer based lightweight network is used to control the model complexity. It integrates fewer model parameters and features higher parameter efficiency.

\section{Experimental Results}
The proposed model was evaluated using Sleep-EDF dataset \cite{kemp2000analysis}, which contains 20 healthy patients with overnight 100Hz-sampled EEG records and corresponding sleep patterns based on AASM.
Only the Fpz-Cz channel was used for training and testing.
The single-channel Fpz-Cz with 20 fold cross-validation was used for evaluation, where both nights from each subject were used. In the test stage, there will be no subjects who have already appeared in the training set.

There are 43.08K parameters in the proposed model. 
The proposed light-weight model prevents the algorithm from over-fitting. In addition, the model can be stored in on-chip SRAM, which greatly reduces the power consumption in reading/writing the memory.
The calculation complexity is 45.76 MFLOPs. The proposed work is implemented on the snapdragon 810 platform. 
It features a less than 1\% occupation of the CPU resources. 
\begin{table*}[!ht]
\caption{Comparison of the ACC ,number of parameters and omputation complexity}
\label{tb:modules}
\begin{center}
\scalebox{0.9}{
\begin{tabular}{|c|c|c|c|c|c|c|c|c|c|c|c|c|c|c|}
\hline
                                                      &                              &                         &                       & {\color[HTML]{000000} }                                      & \multicolumn{2}{c|}{{\color[HTML]{000000} W}}            & \multicolumn{2}{c|}{{\color[HTML]{000000} N1}}          & \multicolumn{2}{c|}{{\color[HTML]{000000} N2}}            & \multicolumn{2}{c|}{{\color[HTML]{000000} N3}}           & \multicolumn{2}{c|}{{\color[HTML]{000000} REM}}          \\ \cline{6-15}
\multirow{-2}{*}{Methods}                             & \multirow{-2}{*}{Parameters} & \multirow{-2}{*}{FLOPs} & \multirow{-2}{*}{ACC} & \multirow{-2}{*}{{\color[HTML]{000000}Test Data}} & {\color[HTML]{000000} TP}   & {\color[HTML]{000000} FN}  & {\color[HTML]{000000} TP}  & {\color[HTML]{000000} FN}  & {\color[HTML]{000000} TP}   & {\color[HTML]{000000} FN}   & {\color[HTML]{000000} TP}   & {\color[HTML]{000000} FN}  & {\color[HTML]{000000} TP}   & {\color[HTML]{000000} FN}  \\ \hline
The proposed Method                                   & 43.08K                       & 45.76M                  & 78.86\%               & {\color[HTML]{000000} 11012}                                 & {\color[HTML]{000000} 1938} & {\color[HTML]{000000} 316} & {\color[HTML]{000000} 198} & {\color[HTML]{000000} 488} & {\color[HTML]{000000} 3495} & {\color[HTML]{000000} 821}  & {\color[HTML]{000000} 1755} & {\color[HTML]{000000} 175} & {\color[HTML]{000000} 1278} & {\color[HTML]{000000} 549} \\ \hline
LightSleepNet+AdaBN                                   & 43.08K                       & 45.76M                  & 78.05\%               & {\color[HTML]{000000} 11012}                                 & {\color[HTML]{000000} 1915} & {\color[HTML]{000000} 339} & {\color[HTML]{000000} 185} & {\color[HTML]{000000} 501} & {\color[HTML]{000000} 3495} & {\color[HTML]{000000} 821}  & {\color[HTML]{000000} 1755} & {\color[HTML]{000000} 174} & {\color[HTML]{000000} 1242} & {\color[HTML]{000000} 585} \\ \hline
LightSleepNet+Gradient Weighting                      & 43.08K                       & 45.76M                  & 77.59\%               & {\color[HTML]{000000} 11012}                                 & {\color[HTML]{000000} 1915} & {\color[HTML]{000000} 339} & {\color[HTML]{000000} 205} & {\color[HTML]{000000} 481} & {\color[HTML]{000000} 3539} & {\color[HTML]{000000} 777}  & {\color[HTML]{000000} 1639} & {\color[HTML]{000000} 290} & {\color[HTML]{000000} 1260} & {\color[HTML]{000000} 567} \\ \hline
LightSleepNet                                         & 43.08K                       & 45.76M                  & 77.42\%               & {\color[HTML]{000000} 11012}                                 & {\color[HTML]{000000} 1848} & {\color[HTML]{000000} 406} & {\color[HTML]{000000} 164} & {\color[HTML]{000000} 522} & {\color[HTML]{000000} 3539} & {\color[HTML]{000000} 777}  & {\color[HTML]{000000} 1716} & {\color[HTML]{000000} 213} & {\color[HTML]{000000} 1242} & {\color[HTML]{000000} 585} \\ \hline
DeepCNN\cite{supratak2017deepsleepnet}                & 614.02K                      & 22.07M                  & 73.36\%               & {\color[HTML]{000000} 11012}                                 & {\color[HTML]{000000} 1645} & {\color[HTML]{000000} 609} & {\color[HTML]{000000} 89}  & {\color[HTML]{000000} 597} & {\color[HTML]{000000} 3150} & {\color[HTML]{000000} 1166} & {\color[HTML]{000000} 1736} & {\color[HTML]{000000} 193} & {\color[HTML]{000000} 1443} & {\color[HTML]{000000} 384} \\ \hline
IGCV1 big\cite{zhang2017interleaved}                  & 234.95K                      & 208.02M                 & 75.70\%               & {\color[HTML]{000000} 11012}                                 & {\color[HTML]{000000} 2028} & {\color[HTML]{000000} 226} & {\color[HTML]{000000} 178} & {\color[HTML]{000000} 508} & {\color[HTML]{000000} 3237} & {\color[HTML]{000000} 1079} & {\color[HTML]{000000} 1736} & {\color[HTML]{000000} 193} & {\color[HTML]{000000} 1151} & {\color[HTML]{000000} 676} \\ \hline
IGCV1 small                                           & 40.13K                       & 34.98M                  & 74.42\%               & {\color[HTML]{000000} 11012}                                 & {\color[HTML]{000000} 1848} & {\color[HTML]{000000} 406} & {\color[HTML]{000000} 240} & {\color[HTML]{000000} 446} & {\color[HTML]{000000} 3150} & {\color[HTML]{000000} 1166} & {\color[HTML]{000000} 1678} & {\color[HTML]{000000} 251} & {\color[HTML]{000000} 1260} & {\color[HTML]{000000} 567} \\ \hline
Super Separable Convolution\cite{kaiser2017depthwise} & 8.26K                        & 8.66M                   & 74.90\%               & {\color[HTML]{000000} 11012}                                 & {\color[HTML]{000000} 1983} & {\color[HTML]{000000} 271} & {\color[HTML]{000000} 150} & {\color[HTML]{000000} 536} & {\color[HTML]{000000} 3323} & {\color[HTML]{000000} 993}  & {\color[HTML]{000000} 1504} & {\color[HTML]{000000} 425} & {\color[HTML]{000000} 1278} & {\color[HTML]{000000} 549} \\ \hline
Time-Distributed Deep CNN\cite{koushik2019real}       & 226.53K                      & 654.72M                 & 73.27\%               & {\color[HTML]{000000} 11012}                                 & {\color[HTML]{000000} 1938} & {\color[HTML]{000000} 316} & {\color[HTML]{000000} 253} & {\color[HTML]{000000} 433} & {\color[HTML]{000000} 2891} & {\color[HTML]{000000} 1425} & {\color[HTML]{000000} 1736} & {\color[HTML]{000000} 193} & {\color[HTML]{000000} 1268} & {\color[HTML]{000000} 559} \\ \hline
\end{tabular}}
\end{center}
\end{table*}

Table \ref{tb:modules} illustrates the comparison between the proposed LightSleepNet and highly-efficient convolution blocks in computer vision in the terms of accuracy, the number of parameters, and the computational complexity.
The block proposed in IGCV1 \cite{zhang2017interleaved}, super separable convolution \cite{kaiser2017depthwise} and Time-Distributed Deep CNN \cite{koushik2019real} was re-implemented for comparison purposes. The experimental results show that the proposed work achieves the best in accuracy with a good trade-off in the number of parameters. As the first and fourth rows in Table \ref{tb:modules} show, the proposed personalized adaptation methodology contributes a 1.44\% improvement for the task without additional parameter and computation cost in inference. \cite{supratak2017deepsleepnet} proposed a classical deep CNN architecture in sleep staging. However, the accuracy performance is worse than most of the light-weight architectures listed in Table \ref{tb:modules} due to the over-fitting with the Sleep-EDF dataset.

The proposed work is also compared with state-of-the-art designs in Table \ref{tb:comparision}.
\cite{sharma2017automatic} features a higher accuracy than the proposed work, but it belongs to the non-independent dataset splitting, in which, data of all subjects has been occurred in the training set, which is equivalent to remove the influence of individual differences and it is not suitable for new subjects who have never seen before.
\cite{phan2019towards} achieves the best performance in single-channel sleep staging. However, it is power hungry while the computational complexity of the proposed work is much lower, which makes the proposed work a better solution for wearable devices. Furthermore, the proposed scheme could be easily applied to EEG signal processing tasks where there is no sufficient data for pre-training.

\begin{table*}[!ht]
\caption{Comparison With The State of The Art Using EEG Fpz-Cz channel}
\label{tb:comparision}
\begin{center}
  \begin{tabular}{|l|l|l|l|l|l|l|l|l|l|l|}
    \hline
    {\color[HTML]{000000} Methods}              & {\color[HTML]{000000} Dataset} & {\color[HTML]{000000} Test Data} & \multicolumn{3}{l|}{{\color[HTML]{000000} Overall Metrics}}                            & \multicolumn{5}{l|}{{\color[HTML]{000000} Per-class F1-Score}}                                                                             \\
    {\color[HTML]{000000} }                     & {\color[HTML]{000000} }        & {\color[HTML]{000000} }          & {\color[HTML]{000000} ACC} & {\color[HTML]{000000} MF1} & {\color[HTML]{000000} kappa} & {\color[HTML]{000000} W}  & {\color[HTML]{000000} N1} & {\color[HTML]{000000} N2} & {\color[HTML]{000000} N3} & {\color[HTML]{000000} REM} \\ \hline
    Proposed Method                             & Sleep-EDF                      & 42308                            & 83.8                       & 75.3                       & 0.78                         & {\color[HTML]{000000} 90} & {\color[HTML]{000000} 31} & {\color[HTML]{000000} 88} & {\color[HTML]{000000} 89} & {\color[HTML]{000000} 78}  \\ \hline
    LightSleepNet Without Personalized Adaptation & Sleep-EDF                      & 42308                            & 83.3                       & 75.3                       & 0.77                         & {\color[HTML]{000000} 90} & {\color[HTML]{000000} 33} & {\color[HTML]{000000} 88} & {\color[HTML]{000000} 89} & {\color[HTML]{000000} 76}  \\ \hline
    IEEE TNSRE17\cite{supratak2017deepsleepnet} & Sleep-EDF                      & 41950                            & 82.0                       & 76.9                       & 0.76                         & {\color[HTML]{000000} 85} & {\color[HTML]{000000} 47} & {\color[HTML]{000000} 86} & {\color[HTML]{000000} 85} & {\color[HTML]{000000} 82}  \\ \hline
    ISCAS20\cite{liao20Trifeature}              & Sleep-EDF                      & 41950                            & 82.9                       & 75.6                       & 0.77 -                       & {\color[HTML]{000000} 90} & {\color[HTML]{000000} 24} & {\color[HTML]{000000} 87} & {\color[HTML]{000000} 95} & {\color[HTML]{000000} 82}  \\ \hline
    Arxiv19 \cite{phan2019towards}              & Sleep-EDF                      & 41950                            & 85.2                       & 79.6                       & 0.79                         & {\color[HTML]{000000} -}  & {\color[HTML]{000000} -}  & {\color[HTML]{000000} -}  & {\color[HTML]{000000} -}  & {\color[HTML]{000000} -}   \\ \hline
    NCA17 \cite{sharma2017automatic}            & Sleep-EDF                      & 15136                            & 91.3                       & 77.0                       & 0.86                         & {\color[HTML]{000000} 98} & {\color[HTML]{000000} 30} & {\color[HTML]{000000} 89} & {\color[HTML]{000000} 86} & {\color[HTML]{000000} 83}  \\ \hline
    IEEE TNSRE18\cite{chambon2018deep}          & Sleep-EDF                      & 37022                            & 81.44                      & 72.2                       & -                            & {\color[HTML]{000000} 81} & {\color[HTML]{000000} 40} & {\color[HTML]{000000} 85} & {\color[HTML]{000000} 76} & {\color[HTML]{000000} 79}  \\ \hline
    EMBC18\cite{phan2018dnn}                    & Sleep-EDF                      & 37022                            & 82.6                       & 74.2                       & 0.76                         & {\color[HTML]{000000} 90} & {\color[HTML]{000000} 33} & {\color[HTML]{000000} 87} & {\color[HTML]{000000} 86} & {\color[HTML]{000000} 75}  \\ \hline
    IEEE TBE 18\cite{phan2018joint}             & Sleep-EDF                      & 37022                            & 81.9                       & 73.8                       & 0.74                         & {\color[HTML]{000000} 76} & {\color[HTML]{000000} 32} & {\color[HTML]{000000} 87} & {\color[HTML]{000000} 87} & {\color[HTML]{000000} 91}  \\ \hline
    BSN 19\cite{koushik2019real}                & Sleep-EDF                      & 41950                            & 83.5                       & -                          & -                            & {\color[HTML]{000000} 89} & {\color[HTML]{000000} 44} & {\color[HTML]{000000} 85} & {\color[HTML]{000000} 86} & {\color[HTML]{000000} 77}  \\ \hline
  \end{tabular}
\end{center}

\end{table*}

\begin{table}[!h]
  \caption{Per-class Metrics For the LightSleepNet}
  \begin{center}
\begin{tabular}{|l|l|l|l|l|}
\hline
{\color[HTML]{000000} Class} & {\color[HTML]{000000} TP}    & {\color[HTML]{000000} FN}   & {\color[HTML]{000000} Sensitivity} & {\color[HTML]{000000} Specificity} \\ \hline
{\color[HTML]{000000} W}     & {\color[HTML]{000000} 7456}  & {\color[HTML]{000000} 829}  & {\color[HTML]{000000} 0.90}        & {\color[HTML]{000000} 0.98}        \\ \hline
{\color[HTML]{000000} N1}    & {\color[HTML]{000000} 644}   & {\color[HTML]{000000} 2160} & {\color[HTML]{000000} 0.23}        & {\color[HTML]{000000} 0.98}        \\ \hline
{\color[HTML]{000000} N2}    & {\color[HTML]{000000} 15663} & {\color[HTML]{000000} 2136} & {\color[HTML]{000000} 0.88}        & {\color[HTML]{000000} 0.91}        \\ \hline
{\color[HTML]{000000} N3}    & {\color[HTML]{000000} 5075}  & {\color[HTML]{000000} 628}  & {\color[HTML]{000000} 0.89}        & {\color[HTML]{000000} 0.98}        \\ \hline
{\color[HTML]{000000} REM}   & {\color[HTML]{000000} 6559}  & {\color[HTML]{000000} 1158} & {\color[HTML]{000000} 0.85}        & {\color[HTML]{000000} 0.93}        \\ \hline
\end{tabular}
\end{center}
\end{table}

\section{Conclusion}
This paper proposed LightSleepNet - a single-channel EEG based, high accuracy personalized sleep staging architecture with high parameter efficiency and low computational complexity. The proposed framework can be implemented on various mobile platforms with limited hardware resources. It achieves a state-of-the-art overall accuracy of 83.8\% with only 45.76 MFLOPs computation and 43.08 K parameters. The latency of the proposed framework is less than 30s for sleep staging with an input of one 30s single-channel EEG epoch. The proposed framework could be personalized for new subject using unlabeled data without re-training, in which the accuracy of LightSleepNet is improved without additional training and computation cost.

\small

\end{document}